\documentclass[aps,pra,reprint,groupedaddress]{revtex4-1} %reprint

\usepackage{amsmath,amsthm,amssymb}
\usepackage{graphicx}

\newcommand{\x}{{\mathbf{r}}}
\renewcommand{\j}{\mathbf{j}}
\newcommand{\A}{\mathbf{A}}

\begin{document}

%\preprint{}

\title{%Non-existence of a weak 
	Density-Wavefunction Mapping 
	in Degenerate Current-Density-Functional Theory}

\author{Andre Laestadius}
\email{andre.laestadius@kjemi.uio.no}
\author{Erik I. Tellgren}
%\email{andre.laestadius@kjemi.uio.no}%Lines break automatically or can be forced with \\
\affiliation{Hylleraas Centre for Quantum Molecular Sciences, Department of Chemistry, University of Oslo, P.O. Box 1033 Blindern, N-0315 Oslo, Norway }
%\author{Erik I Tellgren}%
%\email{michaelb@kth.math.se}
%\affiliation{Department of Mathematics, KTH -- Royal Institute of Technology, 114 28 Stockholm, Sweden}%
\date{\today}% It is always \today, today,


%\author{Andre Laestadius}
%\email[]{andre.laestadius@kjemi.uio.no}
%\thanks{}
%\affiliation{Centre for Theoretical and Computational Chemistry, Department of Chemistry, University of Oslo}


%\date{\today}

\begin{abstract}
We show that the particle density, $\rho(\x)$, and the paramagnetic current density, $\j^{p}(\x)$, are not sufficient 
to determine the set of degenerate ground-state wave functions. 
This is a general feature of degenerate systems where the degenerate states have different angular momenta. We provide a general strategy for constructing Hamiltonians that share the same ground state density, yet differ in degree of degeneracy. We then provide a fully analytical example for a noninteracting system subject to electrostatic potentials and uniform magnetic fields.
Moreover, we prove that when $(\rho,\j^p)$ is ensemble $(v,\A)$-representable by a mixed state formed from $r$ degenerate ground states, then any Hamiltonian $H(v',\A')$ that shares this ground state density pair must have at least $r$ degenerate ground states in common with $H(v,\A)$. Thus, any set of Hamiltonians that shares a ground-state density pair $(\rho,\j^p)$ by necessity has at least have one joint ground state.
\end{abstract}

 % old pace, not updated for this CDFT article

\maketitle

\begin{figure}%[b]
	%\scalebox{.44}{\includegraphics{fig1}}
	\includegraphics[width=0.5\textwidth]{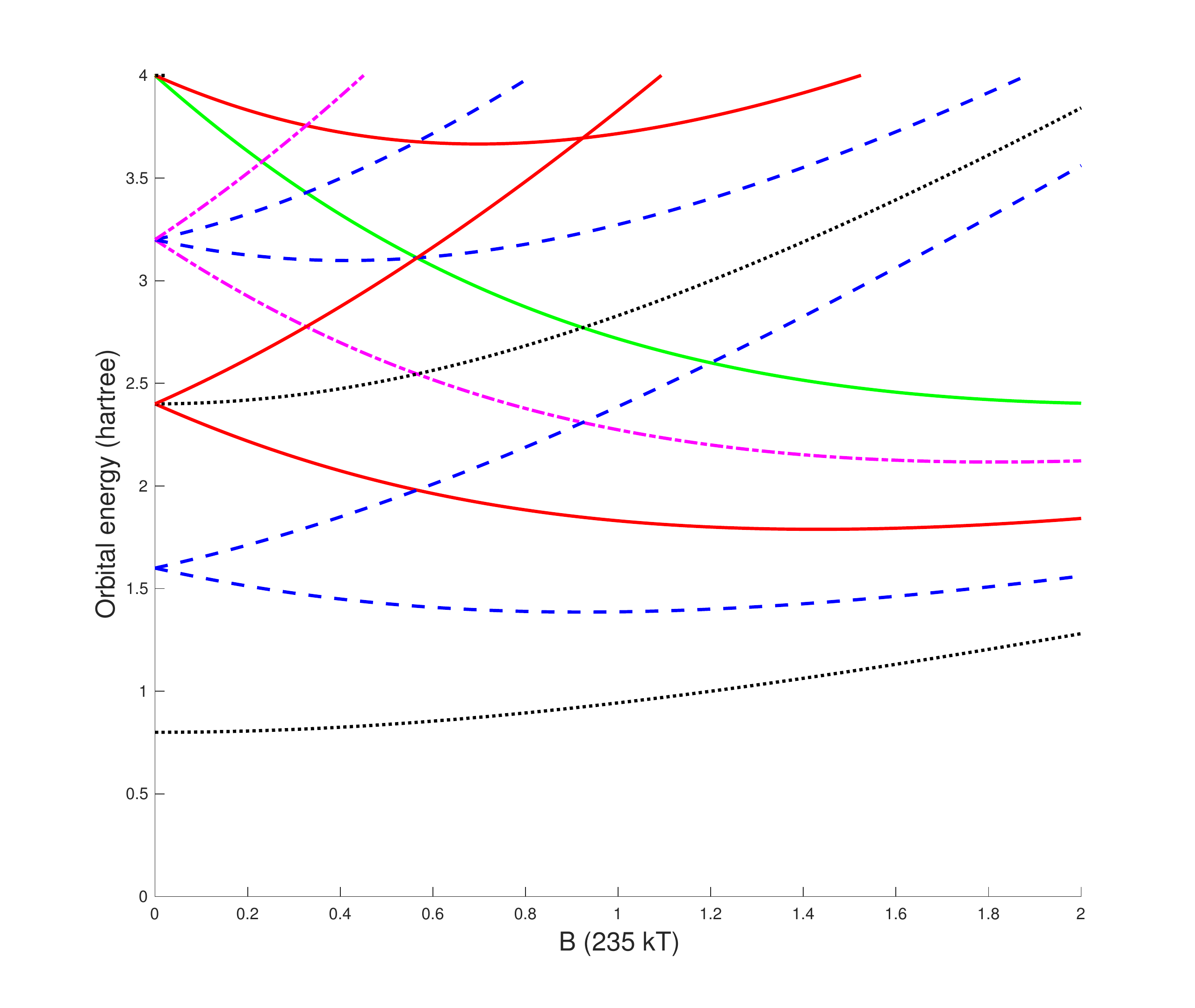}
	\caption{\label{fig:1} Orbital energies $\varepsilon_{n,m}$ (see Eq.~\eqref{eq:Orb}) for $\omega=0.8$ as a function of the magnetic field strength $B\geq 0$. Orbital energies with $m=0$ are shown as dotted black curves, $m=\pm 1$ as dashed blue curves, $m= \pm 2$ as solid red curves, $m = \pm 3$ as dashed dot magenta curves, and $m= \pm 4$ as solid green curves.}
\end{figure}

\begin{figure}%[b]
	%\scalebox{.44}{\includegraphics{fig1}}
	\includegraphics[width=0.5\textwidth]{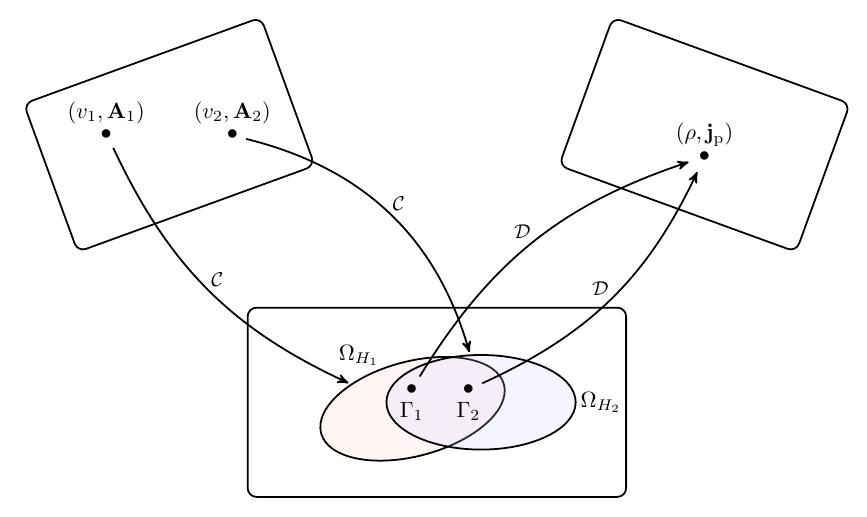}
	\caption{\label{fig:2} Two Hamiltonians have different sets
		of degenerate ground states (indicated by ellipses). Suppose the density matrices
		$\Gamma_1$ and $\Gamma_2$ are ground states of $H(v_1,\A_1)$ and $H(v_2,\A_2)$, respectively. Assume
		further that they map to the same density, $\Gamma_1 \mapsto (\rho, \j_p)$ and $\Gamma_1 \mapsto (\rho,\j_p)$. Then
		it follows that $\Gamma_1$ is also a ground state of $H(v_2,\A_2)$ and that $\Gamma_2$ is also a ground
		state of $H(v_1,\A_1)$. Thus, both $\Gamma_1$ and $\Gamma_2$ are located in the intersection of the two
		ellipses.}
\end{figure}

%\section{Introduction}
A cornerstone of modern density-functional theory (DFT) is the Hohenberg-Kohn theorem \cite{Hohenberg1964}, which states that %knowledge of the one-body density 
the ground-state particle density of a quantum-mechanical system 
determines up to an additive constant the one-body potential $v$ of the same system.
The original argument was limited to systems that have unique ground states. 
Although DFT can be formulated without recourse to the Hohenberg-Kohn theorem, using the constrained-search and Lieb's convex analysis formalisms \cite{Levy,Lieb83}, the result strengthens and adds insight to the theory. Firstly, the alternative DFT formulations establish only that densities determine various contributions to the total energy; in particular, the exchange-correlation energy, and properties given as functional derivatives of the energy with respect to the scalar potential. The stronger statement that the wave function and Hamiltonian, and consequently \emph{all} properties of a system, are determined still requires the Hohenberg-Kohn theorem. Secondly, whereas alternatives are known for ground-state DFT, the available formulation of time-dependent DFT is most closely related to the Hohenberg-Kohn formulation \cite{RungeGross}. \\
\indent When the Hamiltonian contains a magnetic vector potential in addition to the scalar potential, the particle density alone is no longer sufficient for a rigorous formulation of DFT. The most well-established extension is current-density-functional theory (CDFT), where it has been proven that the particle density, $\rho$, and the paramagnetic current density, $\j^p$, determine the non-degenerate ground-state \cite{Vignale1987,Capelle2002} (see Eqs.~\eqref{eq:rho} and \eqref{eq:jp} for the definition of $\rho$ and $\j^p$, respectively). We use the term {\it weak} Hohenberg-Kohn theorem (cf. \cite{Tellgren2017}, Sec. III D) for this result and, following Dreizler and Gross \cite{Dreizler}, denote the invertible map from non-degenerate ground states to densities by $\mathcal D$. The reason for the term {\it weak} is that such a result would be implied by the stronger but false statement that $(\rho,\j^{p})$ determines $(v,\A)$ \cite{Capelle2002,Tellgren2012,Laestadius2014}. Thus, the map, denoted $\mathcal C$, from $(v,\A)$ to non-degenerate ground states is not invertible. The situation can be summarized as 
\begin{equation}
(v,\A) \stackrel{\mathcal C}{\rightarrow} \psi \stackrel{\mathcal D}{\rightleftarrows  } ( \rho, \j^p)~.
\label{eq:map}
\end{equation}
The fact that the map $\mathcal C$ is not invertible does not preclude a density-functional formulation in terms of $\rho$ and $\j^p$. Indeed, the existence of the map $\mathcal D^{-1}$ in non-degenerate paramagnetic CDFT is enough to define a corresponding Hohenberg-Kohn functional \cite{Vignale1987}. 
Furthermore, the Hohenberg-Kohn variational principle holds for the density pair $(\rho,\j^p)$ and a theory of density functionals can be based on these variables \cite{Vignale1987,Capelle2002}. For further discussion on the choice of variables for current-density functionals we refer to \cite{Tellgren2012,Laestadius2014}, see also the mathematical analyses in \cite{Laestadius2014b,Laestadius2014c,Kvaal2015} and the related \cite{Tellgren2017}. For the status of the Hohenberg-Kohn theorem for physical current density instead of the paramagnetic current density, we refer to previous work showing that existing attempted proofs are flawed \cite{Tellgren2012,Laestadius2014} and the recent progress towards a positive result using the total current density \cite{MDFT,RUGGENTHALER}. \newline
\indent The aim of this work is to investigate a weak Hohenberg-Kohn result in CDFT without the assumption of a unique ground state. 
Given an N-electron wave function $\psi$, define the particle density and the paramagnetic current density according to
\begin{align}
\label{eq:rho}
\rho_\psi(\mathbf r_1) &= N \int |\psi|^2\, d\mathbf r_{-1},\\
\j_\psi^p(\mathbf r_1) &= N\,\text {Im} \int \overline{\psi}\,\nabla_{1}\psi  \, d\mathbf r_{-1},
\label{eq:jp}
\end{align}
where $\int d\mathbf r_{-1}$denotes integration over all space for all but one particle and $\overline{\psi}$ denotes the complex conjugate of $\psi$. 
Furthermore, given a vector potential $\A$ we may compute the total current density as the sum $\mathbf j = \mathbf j^p + \rho \A$. \\
\indent For vanishing $\A$, the Hohenberg-Kohn theorem states that if $\rho_1 = \rho_2$, then $V_1 = V_2 \,+$ constant, where $V_k = \sum_j v_k(\mathbf r_j)$ \cite{Hohenberg1964}. 
The proof of this result relies on the fact that if $\psi$ is a ground state of both systems, then 
$(V_1 - V_2)\psi = \text{constant} \times \psi$. If $\psi$ does not vanish on a set of positive (Lebesgue) measure, we have $V_1 =V_2\,+$ constant (almost everywhere). At any rate $V_1= V_2$ up to a constant holds on the complement of $\mathcal N_\psi =\{\psi= 0\}$. 
Assuming that the measure of $\mathcal N_\psi$ is zero (i.e., assuming that that the Schr\"odinger equation has the unique-continuation property from sets of positive measure), the proof can be completed by means of the variational principle as first suggested in \cite{Hohenberg1964}. A generalization of the original Hohenberg-Kohn theorem that includes degeneracy was given in \cite{Englisch1983}. 
[See also the work of Lammert \cite{Lammert2015} for further analysis of the set $\mathcal N_\psi$ in connection with the Hohenberg-Kohn theorem in DFT.] \\
\indent In the presence of a magnetic field, a ground state $\psi$ does not uniquely determine the Hamiltonian $H$ \cite{Capelle2002,Tellgren2012,Laestadius2014}. This leads to complications in the following way:   
We demonstrate that a given pair $\rho$ and $\j^{p}$ may arise from two different pairs 
of $v$ and $\A$ that \emph{do not share the same set of ground states}.
This shows that the conclusion of Theorem~9 in \cite{Laestadius2014} does not hold in general.   
Nonetheless, any set of ground-state density matrices that have the same density pair $(\rho,\j^p)$ are ground states of the same set of Hamiltonians (see also \cite{Capelle2007} and the discussion that comes before Theorem~9 in \cite{Laestadius2014}). We furthermore prove that $(\rho,\mathbf j^p)$ at least determine one ground state, and under certain assumptions, the full set. This constitutes a weak {\it ensemble} Hohenberg-Kohn result in degenerate CDFT.\\
\indent In what follows, our point of departure is a quantum mechanical system of $N$ (spinless) electrons subjected to both a magnetic field and a scalar potential. The 
Hamiltonian is
%magnetic Schr\"odinger operator may be written
\begin{equation*}
%H_{v,\A} 
H(v,\A)= H_0  + \sum_{j=1}^N \Big(\frac 1 2 \{- i \nabla_j,\A(\mathbf r_j)\}   +  v(\mathbf r_j) +\frac 1 2 A(\mathbf r_j)^2   \Big),	
\end{equation*}
where $\{\cdot,\cdot\cdot\}$ denotes the anti commutator and $H_0$ is the universal part of $H$, independent of the external potentials $v$ and $\A$. We let
\begin{equation*}
H_0(\lambda)= \frac 1 2 \sum_{j=1}^N\big( - \nabla_j^2 +  \lambda\sum_{j\neq k} r_{jk}^{-1} \big),\quad 0\leq \lambda \leq 1,
\end{equation*} 
where $\lambda=1$ corresponds to fully interacting electrons and $\lambda=0$ the non-interacting case. \\
%
%
%
%
%
%\indent In the presence of a magnetic field, a general Hohenberg-Kohn result for $(\rho,\j)$ remains to date unproven except the special case $N=1$ \cite{Tellgren2012,Laestadius2014}. For recent progress on this topic, we refer to \cite{MDFT,RUGGENTHALER}.
%\indent \mycomment{In general, the ground-state density pair $(\rho,\mathbf j^p)$ does not determine the potentials $v$ and $\A$, and therefore also not the Hamiltonian} $H$ \cite{Capelle2002,Tellgren2012,Laestadius2014}. Below we 
\indent We start by demonstrating that $(\rho,\mathbf j^p)$ does not determine the set of possibly degenerate ground states. 
The general idea is that for systems with cylindrical symmetry about the $z$-axis, degeneration can either be introduced or lifted by the application of an external magnetic field. For example, consider a cylindrically symmetric Hamiltonian $H(v+A^2/2, \mathbf 0)$ with a ground-state degeneracy, where the ground states are distinguished by different eigenvalues of $L_z$. The Hamiltonian $H(v,\A)$ shares the same eigenstates, but the eigenvalue degeneracies are now lifted by the orbital Zeeman effect. At least for sufficiently weak magnetic fields along the $z$-axis, the state with minimal $L_z$ is then the unique ground state. The idea can also be applied in the other direction. That is, suppose a magnetic field has been tuned so that $H(v,\A)$ has a ground state degeneracy, where the ground states are distinguished by different $L_z$ values. The degeneracy is then lifted in the spectrum of the Hamiltonian $H(v+A^2/2,\mathbf 0)$.\\
%
%
%Old: For a system in a magnetic field with a vector potential $\A$ and a scalar potential $v$, the magnetic field strength is varied until a set of degenerate ground-states is obtained. Crucial for the argument is that one of the ground states, say $\psi_0'$, has $L_z=0$. A different system without magnetic field is then introduced that has $v'= v + A^2/2$.  For this system, $\psi_0'$ is the unique ground state.\\
%
%
\indent In order to avoid relying on numerical results, we shall focus on a two-dimensional non-interacting system of $N$ electrons subject to a magnetic field. 
Define $\mathbf r_j =(x_j,y_j)$, $v(r) =\frac 1 2 \omega^2r^2 $ and $\A = (B/2)(-y,x,0)$, where $B\geq 0$ is the strength of a uniform magnetic field 
perpendicular to the plane, i.e., $\mathbf{B}= B \mathbf e_z$. Since $\{-i\nabla_j, \A(\mathbf r_j)  \}= B L_{z;j}$, the system's Hamiltonian is given by
\begin{align}
H =H_0(\lambda) +  \sum_{j=1}^N \Big(\frac{B}{2} L_{z;j}   + \Big[\frac{B^2}{8}+ \frac{\omega^2}{2} \Big] r_j^2   \Big).
\label{eq:H}
\end{align}
Let $\lambda=0$ such that $H_0 = \sum_{j=1}^N (-\nabla_j^2/2)$. We write $H = \sum_{j=1}^N h_j$, where (dropping the index $j$) the one-electron operator $h$ is given by 
\[
h = -\frac 1 2 \nabla^2  + \frac{B}{2} L_{z}   + \Big[\frac{B^2}{8}+ \frac{\omega^2}{2} \Big] r^2.
\] 
Let $\tilde \omega =\sqrt{(B/2)^2 + \omega^2}$. 
The eigenfunctions of $h$ in polar coordinates fulfill (see for instance \cite{Taut1994})
\begin{align*}
\phi_{n,m}(r,\varphi) & = C r^{|m|} e^{im \varphi} L_n^{|m|}(\tilde \omega r^2) e^{-\tilde \omega r^2/2}, 
\end{align*}
where $L_n^{|m|}$ are the associated Laguerre polynomials, $n=0,1,\dots$ and $m=0,\pm 1 , \dots$ The corresponding eigenvalues, or orbital energies, are given by
\begin{equation}
\varepsilon_{n,m} = (2n+1+|m|)\,\tilde \omega + \frac{mB} 2.
\label{eq:Orb}
\end{equation}
The first few $\varepsilon_{n,m}$ are plotted in Fig.~\ref{fig:1} for a fixed $\omega=\omega_0$.
\indent To prove our claim, Fig.~\ref{fig:1} shows that it is enough to study a system with $N=3$ electrons (other particle numbers are also possible). 
Let $|n,m\rangle$ be the abstract state vector corresponding to the single-particle wave function $\phi_{n,m}$. Set $\omega=\omega_0$, $B =B_0= \omega_0/\sqrt 2 $, $e_0= 15 \omega_0 /\sqrt{8} $ and 
\begin{align*}
\psi_{0}' &= |0,0\rangle \otimes |0,-1\rangle \otimes |0,1\rangle, \\
\psi_0 &= |0,0\rangle \otimes |0,-1\rangle \otimes |0,-2\rangle,
\end{align*}
where $\otimes$ denotes antisymmetrized tensor product. Then by direct computation, using $B_0= \omega_0/\sqrt 2 $,
\begin{align*}
	H \psi_0' &= (\varepsilon_{0,0} + \varepsilon_{0,-1} + \varepsilon_{0,1}) \psi_0' = e_0 \psi_0',\\
	H \psi_0 &= (\varepsilon_{0,0} + \varepsilon_{0,-1} + \varepsilon_{0,-2}) \psi_0 = e_0 \psi_0,
\end{align*} 
and $\psi_0'$ and $\psi_0$ are both degenerate ground states of $H$, with energy $e_0$. 
See also Fig.~\ref{fig:1} where $\varepsilon_{0,1}=\varepsilon_{0,-2}$ at the point $(B_0,7 B_0/2)\approx (0.57,1.98)$ for $\omega_0=0.8$. Furthermore, $L_z \psi_0'=0$ whereas $L_z \psi_0 \neq 0$.\newline
\indent Next, let $H'$ be a Hamiltonian of the form \eqref{eq:H} for a different system of the same number of electrons, but with $B=0$ and $\omega = \sqrt{(B_0/2)^2 + \omega_0^2} $. It then follows from \eqref{eq:H} that
\[
H = H' + \frac{B_0} 2 L_z.
\]
Thus, $\psi_0'$ is the unique ground-state (up to a phase) of $H'$, with energy $e_0$. Note that the given example can be adapted to include spin. Adding the spin-Zeeman term $g\, \mathbf B \cdot \mathbf s/2 $ to the one-electron operator $h$, as well as having each orbital instead doubly occupied, gives a level crossing at a different $B$ and electron number~$N$.\\
\indent Now, if we compute $\rho$ and $\j^{p}$ from just $\psi_0'$, the pair $(\rho,\j^{p})$ is ($v,\A$)-representable from both $H$ and $H'$,
\[
(v,\A) \stackrel{\mathcal C}{\rightarrow}\{\psi_0,\psi_0' \}\rightarrow (\rho,\j^p) \leftarrow \{ \psi_0' \} \stackrel{\mathcal C}{\leftarrow} (v',\mathbf 0)~.
\]
The Hamiltonians $H$ and $H'$ do not share the same set of ground states and consequently, we have proved: \emph{Knowledge of $\rho$ and $\j^{p}$ is not enough to determine the set of ground states}.\\
\indent In order to obtain fully analytical results, we have focused on a non-interacting model system, i.e., $\lambda=0$. 
Another candidate for analytical results is two-electron quantum dot with fully interacting ($\lambda=1$) electrons---we refer to the work in \cite{Taut1994}, see also \cite{Taut2009}---although the fact that exact solutions are only known for a discrete set of parameter values makes this case harder. Furthermore, level-crossings are ubituitous in more complicated systems as well.
% [For the analytical solutions of $\lambda=1$ and $N=2$ we refer to the work in \cite{Taut1994}, see also \cite{Taut2009}.]  %and different electron numbers $N$. Spin singlet and also spin dependent$H$.
%However, level-crossings are ubiquitous in interacting systems as well. 
For example, quantum rings~\cite{Viefers2004}, atomic systems~\cite{AlHujaj2004,Stopkowicz2015}, and molecular systems~\cite{Stopkowicz2015,Lange2012} all feature level crossings of the type analyzed here. The existence of level-crossings does not depend on the presence or absence of the spin-Zeeman term. In particular, the lithium atom in a homogeneous magnetic field exhibits such a level crossing: In \cite{AlHujaj2004} that includes the spin-Zeeman term (see Sec. IV A and Fig.~1, Table~II and Table~III), the ground-state has $L_z=0$ for field strengths up to a certain value after which a level crossing occurs and there are ground states with both $L_z=0$ and $L_z\neq 0$. Arguing as above, we can find a system without a magnetic field that shares the ground-state with $L_z=0$ and furthermore, for this system, the ground state is unique.\\
\indent It is interesting to note that the above situation cannot arise for the hydrogen atom in a uniform magnetic field. Let $H=\frac 1 2 (-i\nabla + \A)^2 - |\x|^{-1}$,  be the Hamiltonian that models a hydrogen atom in a uniform magnetic field generated by the vector potential $\A = \frac B 2 \x_\perp$, $B>0$ and $\x_\perp =(-y,x,0)$. We denote the ground-state energy $e_0$ and let $\lambda_m = \inf_{L_z=m} \mathcal R_{H}$, where $\mathcal R_{H}$ is the Rayleigh-Ritz quotient of $H$, i.e.,  
\[
\lambda_m =\inf_{L_z=m}\frac {\langle \psi, H\psi\rangle}{\langle \psi, \psi\rangle }.
\]
Theorem 4.6 in \cite{Avron1981} states $e_0= \lambda_0$, and furthermore $\lambda_0<\lambda_{-1}< ...$ since $\lim_{|\mathbf r| \to 0+} v=0$. Thus, no level crossing occurs in this system. \\
%
%
% Ett annat sätt att konstruera numeriska exempel på level-crossings är att hålla magnetfältet fixt och variera bindningslängd i t.ex. en molekyler med två atomer.
%
%Om B=0 kan man då hitta exempel "åt andra hållet", dvs. H(u,0) är degenerad, medan H(u-1/2 A^2, A) har unikt grundtillst. Men det blir kanske för lång utläggning.
%
%
%
\indent We now turn to a positive result.  
To obtain a weak ensemble Hohenberg-Kohn result, denote $\Omega_H$ the set of ground states belonging to $H$ and let $\{\psi_k \}_{k=1}^m$ be an orthonormal basis of $\Omega_H$. We here assume that $m<+\infty$, i.e., the multiplicity of the ground-state energy $e_0$ is finite. 
For a basis $\{\psi_k \}_{k=1}^m$, $0\leq \lambda_k\leq 1$ and $\sum_{k=1}^m \lambda_k =1$, let $\Gamma_H (\lambda_1,\dots,\lambda_m) = \sum_{k=1}^m \lambda_k \psi_k\rangle \langle \psi_k$ be a density matrix of $H$. A ground-state particle density $\rho$ and paramagnetic current density $\mathbf j^p$ of $H$ are then given by 
$\rho = \text{Tr}\, \Gamma_H \hat \rho  = \sum_{k=1}^m \lambda_k \rho_{\psi_k}$ and $\mathbf j^p = \text{Tr}\, \Gamma_H \hat{\mathbf j}^p  = \sum_{k=1}^m \lambda_k \mathbf j_{\psi_k}^p$. \newline
\indent Conversely, given a particle density $\rho$ 
and a paramagnetic current density $\mathbf j^p$ 
we say that they are $(v,\A)$-ensemble-representable if there 
exists $H$ with a $\Gamma_H$ such that $\Gamma_H \mapsto (\rho,\mathbf j^p)$. We use the standard shorthand $\Gamma_H \mapsto (\rho,\mathbf j^p)$ to denote 
$\rho = \sum_{k=1}^m \lambda_k \rho_{\psi_k}$ and $\mathbf j^p =\sum_{k=1}^m \lambda_k \mathbf j_{\psi_k}^p$. Here, of course, $\{\psi_k\}_{k=1}^m$ is a basis for $\Omega_H$. \newline
\indent We have: {\it Suppose that $\Gamma_k$ is a ground-state density matrix of $H_k$ and moreover that 
$\Gamma_k\mapsto (\rho,\mathbf j^p)$ for $k=1,2$. Then $\Gamma_1$ is a ground-state density matrix for $H_2$ and vice versa.} \\
\indent We can prove this claim as follows. Writing $H_l = H_k + (H_l-H_k)$, we have for $l\neq k$ 
\begin{align*}
\text{Tr}\, \Gamma_k H_l &= e_k + \int
\mathbf j^p\cdot (\A_l-\A_k) d\mathbf r\\
&\quad  + \int \rho( v_l-v_k+ (A_l^2 - A_k^2)/2 ) d\mathbf r.
\end{align*}
Consequently $\text{Tr}\, \Gamma_1 H_2 + \text{Tr}\, \Gamma_2 H_1=e_1+e_2$. Moreover, since $e_l \leq \text{Tr}\, \Gamma_k H_l$ it follows $e_l= \text{Tr}\, \Gamma_k H_l$ and $\Gamma_k$ is also a ground-state density matrix of $H_l$. The result is illustrated in Fig.~\ref{fig:2}.  \newline
\indent There are some immediate consequences of the above fact. In particular, we stress that a Hohenberg-Kohn functional can still be constructed in the degenerate case, since $F_{HK}(\rho,\j^p)= \text{Tr}\, \Gamma H_0 $ has a unique value independent of which ground state $\Gamma\mapsto (\rho,\j^p)$ that is used. Furthermore, if the ground-states of $H_1$ and $H_2$ are non-degenerate, then 
$\rho_1=\rho_2$ and $\mathbf j_1^p =\mathbf j_2^p$ implies $\Omega_{H_1} = \Omega_{H_2}$. This is the result of Vignale and Rasolt \cite{Vignale1987}. \newline
\indent Returning to the degenerate case, as demonstrated in the first part of this work $\Omega_{H_1} = \Omega_{H_2}$ is not true in general even though 
$\Gamma_k\mapsto (\rho,\mathbf j^p)$. We next introduce a definition.
Given a $(v,\A)$-ensemble-representable density pair $(\rho,\mathbf j^p)$, there exists an $H$ with ground state $\Gamma_H$ such that $\rho =\text{Tr}\, \Gamma_H \hat \rho$ and $\mathbf j^p=\text{Tr}\, \Gamma_H \hat{\mathbf j}^p$. Let $r(\Gamma_H)$ denote the rank of $\Gamma_H$, i.e., the number of nonzero eigenvalues $\lambda_k$ of $\Gamma_H$. 
We have the following weak ensemble Hohenberg-Kohn result: \newline
\indent{\it  Assume that $H_1$ and $H_2$ have the sets of ground-states $\Omega_{H_1}$ 
with (orthonormal) basis $\psi_1,\psi_2,\dots,\psi_m$ and $\Omega_{H_2}$ with (orthonormal) basis 
$\phi_1,\phi_2,\dots,\phi_n$. 
Assume $\Gamma_1\mapsto (\rho_1,\mathbf j_1^p)$ and $\Gamma_2\mapsto(\rho_2,\mathbf j_2^p)$, where % denote the %$(v,\A)$-ensemble-representable 
$\Gamma_k$ is a ground-state density matrix of $H_k$. % and $H_2$, respectively. 
If $\rho_1=\rho_2$ and $\mathbf j_1^p=\mathbf j_2^p$, it 
follows that $\Omega_{H_1}\cap\Omega_{H_2}\neq \emptyset$. Moreover,
with the notation $r_k= r(\Gamma_k)$ %and $r_2= r(\Gamma_2)$, 
then there are at least $max (r_1,r_2)$ linearly independent common ground states of the two systems and $$\text{dim}\,\,\Omega_{H_1}\cap\Omega_{H_2}\geq max(r_1,r_2).$$ If in addition $ r_1=\text{dim}\,\Omega_{H_1}$ and 
$ r_2=\text{dim}\,\Omega_{H_2}$, then $\Omega_{H_1} = \Omega_{H_2}$.} \\ 
\indent To prove the above, assume that $\rho_1=\rho_2=\rho$ and $\mathbf j_1^p=\mathbf j_2^p=\mathbf j^p$. For the first part, suppose 
$\Omega_{H_1}\cap\Omega_{H_2} = \emptyset$ and 
let $\{ \lambda_k\}_{k=1}^m$ satisfy $0\leq \lambda_k \leq 1$ and 
$\sum_k \lambda_k =1$ such that $\rho= \sum_{k=1}^m \lambda_k\rho_{\psi_k}$ and 
$\mathbf j^p=\sum_{k=1}^m \lambda_k \mathbf j_{\psi_k}^p$. We then have strict inequality
\begin{align}
e_2 &< \sum_{k=1}^m \lambda_k \langle \psi_k,H_2 \psi_k \rangle = e_1 -\int \mathbf j^p\cdot (\A_2 - \A_1 )d\mathbf r \nonumber 
\\&\qquad + \int\rho (v_2- v_1 +(A_2^2 - A_1^2)/2)d\mathbf r.
\label{HKmagfieldDEG3}
\end{align}
\indent On the other hand, let $\{ \mu_l\}_{l=1}^n$ satisfy $0\leq \mu_l \leq 1$ and 
$\sum_{l=1}^n \mu_l =1$ such that $\rho= \sum_{l=1}^n \mu_l\rho_{\phi_l}$ 
and $\mathbf j^p=\sum_{l=1}^n \mu_l \mathbf j_{\phi_l}^p$. Again using $\Omega_{H_1}\cap\Omega_{H_2} = \emptyset$, it holds
\begin{align}
e_1 &< \sum_{l=1}^n \mu_l \langle \phi_l,H_1 \phi_l \rangle  
=e_2 - \int \mathbf j^p\cdot(\A_1 -\A_2 )d\mathbf r \nonumber\\
&\qquad + \int \rho (v_1- v_2 + (A_1^2 - A_2^2)/2) d\mathbf r.
\label{HKmagfieldDEG4}
\end{align}
Adding \eqref{HKmagfieldDEG3} and \eqref{HKmagfieldDEG4} gives 
$e_1 + e_2 <e_1 + e_2$, which is a contradiction and $\Omega_{H_1}\cap\Omega_{H_2} \neq \emptyset$.\newline 
\indent For the second part, we use that $\Gamma_k \mapsto (\rho,\j^p)$ implies that $\Gamma_1$ is a ground-state density matrix of $H_2$ (and vice versa). To obtain a contradiction, assume  
$\text{dim}\,\,\Omega_{H_1}\cap\Omega_{H_2} <r_1$. Without loss of generality, let $\psi_1,\psi_2,\dots,\psi_{m'}\in \Omega_{H_2}$ 
and $\psi_{m'+1},\psi_{m'+2},\dots,\psi_{m}\notin \Omega_{H_2}$, where 
$m'< r_1 \leq m$. This implies 
\begin{align*}
\text{Tr}\,\Gamma_1 H_2 = \left( \sum_{k=1}^{m'}   +  \sum_{k=m'+1}^m \right)  \lambda_k  \langle \psi_k,H_2\psi_k \rangle  > e_2
\end{align*}
and $\Gamma_1$ is not a ground-state density matrix of $H_2$. By above, this is a contradiction. Hence, there are at least $r_1$ ground states $\psi_k\in \Omega_{H_2}$. \newline	
\indent The proof that there are at least $r_2$ ground states $\phi_k \in \Omega_{H_1}$ is completely analogous, and we can conclude that there are at least $\max (r_1,r_2)$ common ground states of two systems and 
$\text{dim} \,\,\Omega_{H_1} \cap \Omega_{H_2} \geq \max(r_1,r_2)$. \newline
\indent Lastly, with $r_1=m$ and $ r_2=n$, we obtain from the previous step 
\begin{align*} 
\min(m,n)\geq \text{dim} \,\,\Omega_{H_1} \cap \Omega_{H_2} \geq \max(m,n).
\end{align*}
This can only hold when $m=n$, and consequently $\Omega_{H_1}= \Omega_{H_2}$. This completes the proof. \newline
\indent To summarize, we have proved that a density pair $(\rho,\j^p)$ in general does not determine the full set of ground states. The counterexample we have provided demonstrates that a given $(\rho,\j^p)$ may correspond to either a system with a unique ground state, or a system with degenerate ground states. All that is known is that any system that has $(\rho,\j^p)$ as a ground-state density pair must at least share one ground state. While a fully analytical proof is tractable in special cases, such as noninteracting systems, the counterexample only requires that a level-crossing can be tuned by a magnetic field. Hence, this situation is common and can be established numerically in many systems, such as the lithium atom. %The general idea of the counterexample has further been discussed for the atoms hydrogen and lithium, with different results. 
%
%Moreover, a weak ensemble Hohenberg-Kohn theorem has been formulated and proved, highlighting the complexity of degeneracy in paramagnetic current density-functional theory.
%
Moreover, we have proved a positive result. When $(\rho,\j^p)$ is ensemble $(v,\A)$-representable by a mixed state formed from $r$ degenerate ground states, then any Hamiltonian $H(v',\A')$ that shares this ground state density pair must have at least $r$ degenerate ground states in common with $H(v,\A)$. 
Finally, we emphasize that the complications in CDFT due to degeneracy does not effect the generalized Hohenberg-Kohn functional since any ground-state $\Gamma\mapsto (\rho,\j^p)$ has the same expectation value $\text{Tr}\, \Gamma H_0$. 
%
%
%
%\begin{acknowledgments}

	{\it Acknowledgments.} We acknowledge the support of the Norwegian Research Council through the CoE Hylleraas Centre for Quantum Molecular Sciences Grant No.~262695.
	Furthermore, AL acknowledges support from ERC-STG-2014 Grant Agreement No. 639508 and EIT is grateful for support by the Norwegian Research Council through the Grant No.~240674. 
	We thank A. M. Teale for helpful comments. 
%\end{acknowledgments}
%
%
%

\end{document}